\documentstyle[12pt]{article}

\title{A GENERAL APPROACH FOR CALCULATING COUPLING IMPEDANCES
 OF SMALL DISCONTINUITIES }
\author{Sergey~S.~Kurennoy and Robert~L.~Gluckstern \\
        Physics Department, University of Maryland,
        College Park, MD 20742, USA \\
    and Gennady~V.~Stupakov \\ SLAC, Stanford, CA 94309, USA }

\begin{document}

\maketitle
\thispagestyle{empty}\pagestyle{empty}

\begin{abstract}
A general theory of the beam interaction with small discontinuities
of the vacuum chamber is developed taking into account the reaction
of radiated waves back on the discontinuity. The reactive impedance
calculated earlier is reproduced as the first order, and the resistive
one as the second order of a perturbation theory based on this general
approach. The theory also gives, in a very natural way, the results
for the trapped modes due to small discontinuities obtained earlier
by a different method.
\end{abstract}

\section{Introduction}

A common tendency in design of modern accelerators is to minimize
beam-chamber coupling impedances to avoid beam instabilities and
reduce heating. Even contributions from tiny discontinuities
like pumping holes have to be accounted for, due to their large
number, which makes analytical methods for calculating the
impedances of small discontinuities very important. According to
the Bethe theory of diffraction by small holes \cite{Bethe},
the fields diffracted by a hole can be found as those radiated by
effective electric and magnetic dipoles. The coupling impedance of
pumping holes in the vacuum chamber walls has been calculated earlier
\cite{SK,RLG,GS} using this idea. The imaginary part of the impedance
is proportional to the difference of hole polarizabilities
$(\psi -\chi)$, where the magnetic susceptibility $\psi$ and the
electric polarizability $\chi$ are small compared to the cubed typical
dimension $b^3$ of the chamber cross section. From considerations of
the energy radiated into the chamber and through the hole, the real
part of the hole impedance comes out to be proportional to
$(\psi^2 +\chi^2)$, being usually much smaller than the reactance.

In the present paper we further develop this analytical approach by
taking into account the reaction of radiated waves back on the
discontinuity. It leads to a more general theory, and allows us to
reproduce easily all previous results, including those about trapped
modes due to small discontinuities \cite{S&K}. While our consideration
here is restricted to small holes, it can be readily
applied to other small discontinuities like enlargements or irises
because the idea of effective polarizabilities works equally well
in these cases also \cite{K&S}.

\section{General Analysis}

Let us consider an infinite cylindrical pipe with an arbitrary
cross section $S$ and perfectly conducting walls. The $z$ axis is
directed along the pipe axis, a hole is located at the point
($\vec{b},z=0$), and a typical hole size $h$ satisfies $h\ll b$.
To evaluate the coupling impedance one has to calculate the fields
induced in the chamber by a given current.
If an ultrarelativistic point charge $q$ moves along the chamber
axis, the fields harmonics $\vec{E}^b,\vec{H}^b$ produced by this
charge on the chamber wall without hole would be
\begin{equation}
 E^b_{\nu} (z;\omega ) = Z_0H^b_{\tau}  =  -Z_0qe^{ikz}
\sum_{n,m} \frac{e_{nm}(0) \nabla _{\nu}e_{nm}(\vec{b})}
 {k^{2}_{nm}} \ ,        \label{beamf}
\end{equation}
where $k^2_{nm}$, $e_{nm}(\vec{r})$ are eigenvalues and
orthonormalized eigenfunctions (EFs) of the 2D boundary problem
in $S$:
\begin{equation}
\left (\nabla ^2+ k^2_{nm}\right ) e_{nm} =
   0 \ ; \; \; e_{nm}\big\vert_{\partial S} = 0 \ .  \label{boundpr}
\end{equation}
Here $\vec{\nabla}$ is the 2D gradient in plane $S$; $k=\omega/c$;
$\hat{\nu}$ means an outward normal unit vector, $\hat{\tau}$ is
a unit vector tangent to the boundary $\partial S$ of the chamber
cross section $S$, and $\{ \hat{\nu},\hat{\tau},\hat{z}\}$ form a
RHS basis.

\subsection{Fields}

At distances $l$ such that $h \ll l \ll b$, the fields radiated by the
hole into the pipe are equal to those produced by effective dipoles
\cite{Bethe,Collin}\footnote{Polarizabilities $\psi, \chi$ are related
to the effective ones $\alpha_e, \alpha_m$ used in \cite{Collin,SK} as
$\alpha_e=-\chi/2$ and $\alpha_m=\psi/2$, so that for a circular hole
of radius $a$ in a thin wall $\psi=8a^3/3$ and $\chi=4a^3/3$.}
\begin{eqnarray}
P_\nu & = & - \chi \varepsilon_0 E^h_\nu/2; \quad
M_\tau = (\psi_{\tau \tau} H^h_\tau + \psi_{\tau z} H^h_z )/ 2;
 \nonumber \\
M_z & = & (\psi_{z \tau} H^h_\tau + \psi_{z z} H^h_z )/ 2
 \ ,                 \label{dip}
\end{eqnarray}
where superscript '$h$' means that the fields are taken at the hole.
In general, $\psi$ is a symmetric 2D-tensor. If the hole is symmetric,
and its symmetry axis is parallel to $\hat{z}$, the skew terms
vanish, i.e.\ $\psi_{\tau z}=\psi_{z\tau}=0$.

When the effective dipoles are obtained, e.g., by substituting beam
fields (\ref{beamf}) into Eqs.~(\ref{dip}), one can calculate the fields
in the chamber as a sum of waveguide eigenmodes excited in the chamber
by the dipoles, and find the impedance. This approach has been carried
out in \cite{SK}, and for an arbitrary chamber in \cite{SK92}.
However, a more refined theory should take into account the reaction
of radiated waves back on the hole. The TM-eigenmodes contribution
to the radiated fields is a series
\begin{equation}
\vec{F} = \sum_{nm}\left [ A^{+}_{nm}\vec{F}^{+}_{nm}\theta (z) +
    A^{-}_{nm}\vec{F}^{-}_{nm}\theta (-z) \right ] \ ,  \label{fexp}
\end{equation}
where $\vec F$ means either $\vec E$ or $\vec H$ and superscripts
'$\pm$' denote waves radiated respectively in the positive (+, $z>0$)
or negative ($-$, $z<0$) direction. The fields of $\{n,m\}$th
TM-eigenmode in Eq.~(\ref{fexp}) are expressed \cite{Collin} in terms
of EFs (\ref{boundpr})
\begin{eqnarray}
E^\mp_z & = & k_{nm}^2 e_{nm} \exp(\pm \Gamma _{nm}z) \ ;
     \qquad  H^\mp_z = 0 \ ; \nonumber \\
\vec{E}^\mp_t & = & \pm \Gamma _{nm} \vec{\nabla}e_{nm}
        \exp(\pm \Gamma _{nm}z) \ ;               \label{emode} \\
\vec{H}^\mp_t & = & \frac{ik}{Z_0} \hat{z} \times \vec{\nabla}
   e_{nm} \exp(\pm \Gamma _{nm}z) \ , \nonumber
\end{eqnarray}
where propagation factors $\Gamma _{nm}=(k_{nm}^2-k^2)^{1/2}$ should
be replaced by $-i \beta _{nm}$ with $\beta _{nm}=(k^2-k_{nm}^2)^{1/2}$
for $k>k_{nm}$. For given values of dipoles (\ref{dip}) the unknown
coefficients $A^\pm_{nm}$ can be found \cite{SK92} using the Lorentz
reciprocity theorem
\begin{equation}
A^{\pm}_{nm} = a_{nm} M_\tau \pm b_{nm} P_\nu \ ,  \label{Apm}
\end{equation}
\begin{equation}
 a_{nm}  = - \frac{ik Z_0}{2 \Gamma _{nm} k_{nm}^2}
                \nabla_\nu e^h_{nm} \ ;  \quad
 b_{nm} =  \frac{1}{2 \varepsilon_0 k_{nm}^2}
                \nabla_\nu e^h_{nm} \ .  \label{ab}
\end{equation}
In a similar way, the contribution of TE-eigenmodes to the radiated
fields is given by an analogue of Eq.~(\ref{fexp}) with the excitation
coefficients
\begin{equation}
B^{\pm}_{nm} = \pm c_{nm} M_\tau + d_{nm} P_\nu
                              +  q_{nm} M_z  \ ,  \label{Bpm}
\end{equation}
\begin{eqnarray}
 c_{nm} & = & \frac{1}{2 k'^2_{nm}} \nabla_\tau h^h_{nm} \ ;
 \quad q_{nm}  =  \frac{1}{2 \Gamma'_{nm} } h^h_{nm} \ ;
 \nonumber  \\
 d_{nm} & = & - \frac{ik }{2 Z_0 \varepsilon_0
 \Gamma'_{nm} k'^2_{nm}} \nabla_\tau h^h_{nm} \ ,        \label{cdq}
\end{eqnarray}
where EFs $h_{nm}$ satisfy the boundary problem (\ref{boundpr}) with
the Neumann boundary condition $\nabla_\nu h_{nm}\vert_{\partial S}
= 0$, and $k'^2_{nm}$ are corresponding eigenvalues.

Now we can add corrections to the beam fields (\ref{beamf}) due to
the radiated waves in the vicinity of the hole. It gives
\begin{eqnarray}
E_\nu & = & \frac{ E^b_\nu + \psi_{z\tau} \Sigma'_x Z_0 H_\tau +
 \psi_{zz} \Sigma'_x Z_0 H_z } { 1 - \chi ( \Sigma_1 - \Sigma'_1 ) } ,
                                                      \label{En} \\
H_\tau & = & \frac{H^b_\tau + \psi_{\tau z} ( \Sigma_2 - \Sigma'_2 )
 H_z} {1 -\psi_{\tau\tau} (\Sigma_2 - \Sigma'_2 )} ,  \label{Ht} \\
H_z & = & \frac { \chi \Sigma'_x E_\nu  /Z_0 + \psi_{z\tau}
 \Sigma'_3 H_\tau } { 1 - \psi_{zz} \Sigma'_3 }  ,    \label{Hz}
\end{eqnarray}
where ($s=\{n,m\}$ is a generalized index)
\begin{eqnarray}
\Sigma_1 & = & \frac{1}{4}
 \sum_s { \frac{ \Gamma_s \left (\nabla_\nu e^h_s \right)^2}
  {  k^2_s } }  \ ;  \
\Sigma_2  =  \frac{k^2}{4}
 \sum_s { \frac{\left (\nabla_\nu e^h_s \right)^2}
  { \Gamma_s k^2_s } }  \ ;               \nonumber  \\
\Sigma'_1 & = &  \frac{k^2}{4}
 \sum_s \frac{ \left (\nabla_\tau h^h_s \right)^2}
       {\Gamma'_s k'^2_s}  \ ; \
\Sigma'_2  =  \frac{1}{4}
 \sum_s \frac{\Gamma'_s \left (\nabla_\tau h^h_s \right)^2}
       {k'^2_s}  \ ;              \nonumber   \\
\Sigma'_x & = &  i \frac{k}{4} \sum_s
\frac{ h^h_s \nabla_\tau h^h_s }{ \Gamma'_s }  \ ; \
\Sigma'_3  =  \frac{1}{4}
 \sum_s \frac{ k'^2_s \left ( h^h_s \right)^2}
       {\Gamma'_s }  \ .                   \label{sums}
\end{eqnarray}
Since this consideration works at distances not shorter than $l$,
and $l > h$, the summation in Eq.~(\ref{sums}) should be restricted
to values of $s=\{n,m\}$ such that $k_s h \le 1$ and $k'_s h \le 1$.

\subsection{Impedance}

The longitudinal impedance of the hole is defined as
\begin{equation}
Z(k) = - \frac{1}{q} \int _{-\infty}^{\infty}
dz e^{-ikz} E_z(0,z;\omega ) \ ,          \label{impdef}
\end{equation}
where the field at the axis is given by Eq.~(\ref{fexp}) with
coefficients (\ref{Apm}) and (\ref{Bpm}) in which the corrected
near-hole fields (\ref{En})-(\ref{Hz}) are substituted. It yields
\begin{eqnarray}
Z(k) & = & - \frac{ikZ_0 {\tilde{e}_\nu}^2 } {2}
 \left [ \ \frac{\psi_{\tau\tau}}{1 - \psi_{\tau\tau}
( \Sigma_2 - \Sigma'_2 )}\right.                \label{imp} \\
 & &   \left.  + \ \psi^2_{\tau z} \Sigma'_3 - \frac{\chi}
  {1 - \chi ( \Sigma_1 - \Sigma'_1 )} \ \right ] \ , \nonumber
\end{eqnarray}
where $\tilde{e}_\nu \equiv E^b_\nu/(Z_0q) = - \sum_s e_s(0)
\nabla _{\nu}e_s(\vec{b})/k^2_s$ is merely the normalized electric
field produced at the hole location by the beam moving along the
chamber axis, cf.\ Eq.~(\ref{beamf}). In deriving this result we
have neglected the coupling terms between $E_\nu$, $H_\tau$ and
$H_z$, cf.\ Eqs.~(\ref{En})-(\ref{Hz}), which contribute to the
third order of an expansion discussed below, and also have taken
into account that $\psi_{\tau z}=\psi_{z \tau }$.

For a small discontinuity, polarizabilities $\psi, \chi = O (h^3)$,
and they are small compared to $b^3$. If we expand the impedance
(\ref{imp}) in a perturbation series in polarizabilities, the first
order gives
\begin{equation}
Z_1(k) = -\frac{ikZ_0 {\tilde{e}_\nu}^2 } {2}
 \left ( \psi_{\tau\tau}  -  \chi \right ) \ ,          \label{Z1}
\end{equation}
that is exactly the inductive impedance obtained in \cite{SK92} for
an arbitrary cross section of the chamber. For a particular case of
a circular pipe,  from either direct summation in
(\ref{beamf}) or applying the Gauss law, we get $\tilde{e}_\nu =
1/(2\pi b)$, substitution of which into Eq.~(\ref{Z1}) leads to
a well-known result \cite{SK,RLG}. From a physical point of view,
keeping only the first order term (\ref{Z1}) corresponds to
dropping out all radiation corrections in Eqs.~(\ref{En})-(\ref{Hz}).

These corrections first reveal themselves in the second order term
\begin{eqnarray}
Z_2(k)  = -\frac{ikZ_0 {\tilde{e}_\nu}^2 } {2} \left [
 \psi^2_{\tau\tau} ( \Sigma_2 - \Sigma'_2 ) +
 \psi^2_{\tau z} \Sigma'_3   \right.     \\   \label{Z2}
 \left. + \ \chi^2 (\Sigma'_1 - \Sigma_1) \ \right ] \nonumber \ ,
\end{eqnarray}
which at frequencies above the chamber cutoff has both a real
and imaginary part. The real part of the impedance is
\begin{eqnarray}
Re Z_2(k)  & = & \frac{k^3Z_0 {\tilde{e}_\nu}^2 } {8}
 \left \{  \psi^2_{\tau z} \sum^{<}_s \frac{k'^2_s
 \left (h^h_s \right)^2}{k^2 \beta'_s  }  \right. \label{ReZ} \\
 & + & \left. \psi^2_{\tau\tau}  \left [ \sum^{<}_s
   \frac{\left (\nabla_\nu e^h_s \right)^2} { \beta_s k^2_s }
 + \sum^{<}_s \frac{\beta'_s \left (\nabla_\tau h^h_s \right)^2}
       {k^2 k'^2_s}  \right ] \right.    \nonumber  \\
 & + & \left. \chi^2 \left [ \sum^{<}_s \frac{\beta_s
     \left (\nabla_\nu e^h_s \right)^2} {k^2 k^2_s}
 +  \sum^{<}_s \frac{ \left (\nabla_\tau h^h_s \right)^2}
       {\beta'_s k'^2_s} \right ] \right \}  \ , \nonumber
\end{eqnarray}
where the sums include only a finite number of the eigenmodes
propagating in the chamber at a given frequency, i.e.\ those
with $k_s<k$ or $k'_s<k$.

The real part of the impedance is related to the power $P$
scattered by the hole into the beam pipe, $Re\,Z=2P/q^2$, and can
be calculated in an alternative way from energy considerations:
$ P = \sum_s (A^2_sP^{(E)}_s + B^2_sP^{(H)}_s ) $, where we sum over
all propagating modes in both directions, and $P_s$ means the
time-averaged power radiated in $s$th eigenmode:
$P^{(E)}_s=k\beta_sk^2_s/(2Z_0)$ and $P^{(H)}_s=Z_0k\beta'_sk'^2_s/2$.
Substituting beam fields (\ref{beamf}) into
Eqs.~(\ref{Apm})-(\ref{cdq}) for the coefficients $A_s$ and $B_s$
and performing calculations gives the result (\ref{ReZ}).
Such an alternative derivation of the real part has been first
carried out in Ref.~\cite{GS} for a circular pipe with a symmetric
untilted hole ($\psi_{\tau z}=0$). The result (\ref{ReZ}) for this
particular case coincides with that of \cite{GS}. Moreover, in this
case at high frequencies the series can be summed approximately
 \cite{GS} to give $Re\,Z = Z_0 k^4 {\tilde{e}_\nu}^2
(\psi^2_{\tau\tau} + \chi^2)/(12 \pi )$, which can also be obtained
by calculating the energy radiated by the dipoles in a half-space
 \cite{SK92}. Note that the additional $\psi^2_{\tau z}$-term in
Eq.~(\ref{ReZ}) could give a leading contribution to $Re\,Z$, e.g.,
for a long and slightly tilted slot.

\subsection{Trapped Modes}

So far we considered the perturbation expansion of Eq.~(\ref{imp})
implicitly assuming that correction terms $O(\psi)$ and $O(\chi)$
in the denominators of its RHS are small compared to 1.
Under certain conditions this assumption is incorrect,
and it leads to some non-perturbative results. Indeed,
at frequencies slightly below the chamber cut-offs, $0 <  k_s - k
 \ll k_s$, --- or the same with replacement $k_s \to k'_s$, ---
a single term in sums $\Sigma'_1$, $\Sigma_2$, or $\Sigma'_3$ becomes
very large, due to very small $\Gamma_s=(k_s^2-k^2)^{1/2}$ (or
$\Gamma'_s$) in its denominator, and then the ``corrections''
$\psi \Sigma$ or $\chi \Sigma$ can be of the order of 1. As a result,
one of the denominators of the RHS of Eqs.~(\ref{imp}) can vanish,
which corresponds to a resonance of the coupling impedance.
On the other hand, vanishing denominators in
Eqs.~(\ref{En})-(\ref{Hz}) mean the existence of
non-perturbative eigenmodes of the chamber with a hole, since
non-trivial solutions $E,H \neq 0$ exist even for vanishing external
(beam) fields $E^b,H^b = 0$. These eigenmodes are nothing but
the trapped modes studied in \cite{S&K} for a circular waveguide
with a small discontinuity (see \cite{K&S95} for waveguides with
an arbitrary cross section).

Let us for brevity restrict ourselves to the case $\psi_{\tau z}=0$
and consider Eq.~(\ref{Ht}) in more detail. For $H^b = 0$ we
have
\begin{equation}
H_\tau \left [ 1 - \psi_{\tau\tau} \frac { k^2 \left
 (\nabla_\nu e^h_{nm} \right)^2} {4 \Gamma_{nm} k^2_{nm} }
 + \ldots \right ] = 0 \ ,             \label{Htm}
\end{equation}
where $\ldots$ means all other terms of the series $\Sigma_2,
\Sigma'_2$. At frequency $k \simeq k_{nm}$ slightly below the
cutoff $k_{nm}$ of the TM${}_{nm}$-mode, the fraction in
Eq.~(\ref{Htm}) is large due to small $\Gamma_{nm}$ in its
denominator, and one can neglect the other terms. Then the
condition for a non-trivial solution $H_\tau \neq 0$ to exist is
\begin{equation}
\Gamma_{nm} \simeq \frac {1}{4} \psi_{\tau\tau} \left
 (\nabla_\nu e^h_{nm} \right)^2 \ .                \label{GamE}
\end{equation}
This equation gives us the frequency shift $\Delta f$ of the
trapped TM-mode down from the cutoff $f^{(E)}_{nm}$
\begin{equation}
 \frac{\Delta f}{f^{(E)}_{nm}} \simeq \frac{1}{32 k^2_{nm}}
  \psi^2_{\tau\tau} \left (\nabla_\nu e^h_{nm} \right)^4
  \ .  \label{dfE}
\end{equation}

One can easily see that denominator $[1 - \chi ( \Sigma_1 -
\Sigma'_1 )]$ in Eq.~(\ref{En}) does not vanish because singular
terms in $\Sigma'_1$ have a ``wrong'' sign. However, due to the
coupling between $E_\nu$ and $H_z$, a non-trivial solution
$E_\nu,H_z \neq 0$ of simultaneous equations (\ref{En})
and (\ref{Hz}) can exist, even when $E^b = 0$. The corresponding
condition has the form
\begin{equation}
\Gamma'_{nm} \simeq \frac{1}{4} \left [ \psi_{zz} k'^2_{nm}
 \left ( h^h_{nm} \right)^2 - \chi \left
 (\nabla_\tau h^h_{nm} \right)^2 \right ] \ ,      \label{GamH}
\end{equation}
which gives the frequency of the trapped TE${}_{nm}$-mode.

One can easily show that for the particular case of a circular
pipe the results (\ref{GamE})-(\ref{GamH}) coincide with those
obtained by a different method in Ref.~\cite{S&K}. For more
detail, a physical picture of, and resonance impedances due to
trapped modes, see \cite{S&K} and \cite{K&S95}.

\section{Conclusions}

The analytical approach discussed above provides a general
picture for the coupling impedance of a small discontinuity
of the vacuum chamber. It gives the real and imaginary part of the
impedance, as well as trapped modes. Results for typical shapes of
the chamber cross section (circular or rectangular) are easily
obtained from the expressions above using specific EFs, see, e.g.,
in \cite{Collin} or \cite{K&S95}. The transverse impedance can be
derived in a similar way \cite{SK92}.

We have not considered explicitly effects of the wall thickness,
assuming that the hole polarizabilities are the inside ones \cite{RLG},
and they include these effects. We also neglected the radiation
escaping through the hole, contributions of which to the real part
of the impedance are estimated \cite{SK,RLG,S&K}, and usually are
small.

At high frequencies (near or above the chamber cutoff) the mutual
interaction of many holes is important and can cause resonances
if the hole pattern is periodic, e.g.\ \cite{SK93,GS}. A more
complete theory should take this interaction into account.

\end{document}